\documentclass[useAMS,usenatbib]{mn2e}

\usepackage{epsfig}
\usepackage{amsmath}
\usepackage{amssymb}
\usepackage{natbib}
\usepackage{threeparttable} 
\usepackage{booktabs}

\usepackage{epstopdf}
\usepackage{times}

\newcommand{\chan}{\textit{Chandra}}
\newcommand{\swift}{\textit{Swift}}

\newcommand{\Msun}{\mathrm{M}_{\odot}}
\newcommand{\lum}{\mathrm{erg~s}^{-1}}
\newcommand{\flux}{\mathrm{erg~cm}^{-2}~\mathrm{s}^{-1}}

\newcommand{\cnts}{\mathrm{c~s}^{-1}}
\newcommand{\mdot}{\mathrm{M_{\odot}~yr}^{-1}}
\newcommand{\mdotgs}{\mathrm{g~s}^{-1}}
\newcommand{\nh}{\mathrm{cm}^{-2}}
\newcommand{\dist}{(D/5.5~\mathrm{kpc})^2}

\newcommand{\source}{Swift~J174805.3--244637}

\newcommand{\exo}{EXO 0748--676}
\newcommand{\igr}{IGR J17480--2446}

\newcommand{\sax}{SAX J1808.4--3658}
\newcommand{\hete}{HETE J1900.1--2455}

\newcommand{\xte}{XTE J1701--462}

\newcommand{\ks}{KS~1731--260}
\newcommand{\mxb}{MXB~1659--29}
\newcommand{\rx}{XTE~J1709--267}
\newcommand{\maxisource}{MAXI~J0556--332}

\newcommand{\exoter}{EXO 1745--248}
\newcommand{\igrmontse}{IGR J17494--3030}

\def \mnras {MNRAS}
\def \apj {ApJ}
\def \apjs {ApJS}
\def \apjl {ApJL}
\def \aap {A\&A}

\def \atel {Astron. Telegrams}

\def \pasj {PASJ}

\def \prc {Phys. Rev. C}

\title[Crust cooling in \source]{Neutron star crust cooling in the Terzan 5 X-ray transient\\ \source}
\author[N. Degenaar et al.]
{N. Degenaar$^1$\thanks{e-mail: degenaar@ast.cam.ac.uk}, R. Wijnands$^{2}$, A.~Bahramian$^3$, G.R.~Sivakoff$^3$, C.O.~Heinke$^3$\thanks{Alexander von Humboldt Fellow at the Max-Planck-Institut f\"{u}r Radioastronomie, Auf dem H\"{u}gel 69, D-53121 Bonn, Germany}, E.F. Brown$^{4}$,
\newauthor J.K.~Fridriksson$^2$, J.~Homan$^{5,6}$, E.M.~Cackett$^7$, A. Cumming$^{8}$, J.M.~Miller$^{9}$, D.~Altamirano$^{10}$
\newauthor and D.~Pooley$^{11,12}$\\ \\
$^1$Institute of Astronomy, University of Cambridge, Madingley Road, Cambridge CB3 OHA, UK\\
$^2$Anton Pannekoek Institute for Astronomy, University of Amsterdam, Science Park 904, 1098 XH, Amsterdam, the Netherlands\\
$^3$Department of Physics, University of Alberta, 4-183 CCIS, Edmonton, AB T6G 2E1, Canada\\
$^4$Department of Physics and Astronomy, Michigan State University, East Lansing, MI 48824, USA\\
$^5$Massachusetts Institute of Technology, Kavli Institute for Astrophysics and Space Research, Cambridge, MA 02139, USA\\
$^6$SRON, Netherlands Institute for Space Research, Sorbonnelaan 2, 3584 CA, Utrecht, The Netherlands\\
$^7$Department of Physics and Astronomy, Wayne State University, 666 W. Hancock St, Detroit, MI 48201, USA\\
$^8$Department of Physics, McGill University, 3600 rue University, Montreal, QC, H3A 2T8, Canada\\
$^9$Department of Astronomy, University of Michigan, 1085 South University Avenue, Ann Arbor, MI  48109, USA\\
$^{10}$School of Physics and Astronomy, University of Southampton, Southampton, Hampshire, SO171BJ, UK\\
$^{11}$Department of Physics, Sam Houston State University, Huntsville, TX 77341, USA\\
$^{12}$Eureka Scientific, Inc., 2452 Delmer Street, Suite 100, Oakland, CA 94602, USA
}

\begin{document}

\date{Accepted 2015 May 6. Received 2015 April 23.}

\pagerange{\pageref{firstpage}--\pageref{lastpage}} \pubyear{0000}

\maketitle

\label{firstpage}

\begin{abstract}
When neutron stars reside in transient X-ray binaries, their crustal layers become heated during accretion outbursts and subsequently cool in quiescence. Observing and modelling this thermal response has yielded valuable insight into the physics of neutron star crusts. However, one  unresolved problem is the evidence in several neutron stars for an extra energy source, located at shallow depth in the crust, that is not accounted for by standard heating models. Its origin remains puzzling, and it is currently unclear whether this additional heating occurs in all neutron stars, and if the magnitude is always the same. Here, we report on \chan\ observations that cover two years after the 2012 outburst of the transient neutron star X-ray binary \source\ in the globular cluster Terzan 5. The temperature of the neutron star was elevated during the first two months following its $\simeq$8~week accretion episode, but had decayed to the pre-outburst level within $\simeq$100~d. Interpreting this as rapid cooling of the accretion-heated crust, we model the observed temperature curve with a thermal evolution code. We find that there is no need to invoke shallow heating for this neutron star, although an extra energy release up to $\simeq$1.4~MeV~nucleon$^{-1}$ is allowed by the current data (2$\sigma$ confidence). We also present two new data points on the crust-cooling curve of the 11-Hz X-ray pulsar \igr\ in Terzan 5, which was active in 2010. The temperature of this neutron star remains significantly above the pre-outburst level, but we detect no change in the thermal emission since the previous measurements of 2013 February. This is consistent with the slow crust cooling expected several years post-outburst.
\end{abstract}


%%%%%%%%%%%%%%%%%
% INTRODUCTION
%%%%%%%%%%%%%%%%%

\begin{table*}
\caption{Overview of neutron star LMXBs for which crustal cooling has been monitored.}
\begin{threeparttable}
\begin{tabular*}{1.01\textwidth}{@{\extracolsep{\fill}} l c c c c c c c c c}
\hline 
Source name & $t_0$  & $t_{\mathrm{ob}}$ & $\dot{M}_{\mathrm{ob}}$ & $D$ [ref] & $\nu_{\mathrm{spin}}$ [ref] & $P_{\mathrm{orb}}$ [ref] & $kT^{\infty}_{\mathrm{base}}$ & Remarks & Cooling  \\
& (MJD) &  (yr) & ($\mdotgs$) & (kpc) & (Hz) & (hr) & (eV) & & references \\
\hline
\ks & 51930.5 & 12.3 & $1.9\times10^{17}$ & 7.0 [1] & 524 [2] & ... & ... & Superburst source & 3,4,5,6\\
\mxb & 52159.5 & 2.4 & $1.1\times10^{17}$ & 10 [7] & 567 [8] & 7.1 [9] & ... & Eclipsing & 5,10,11,12,13\\
\xte & 54322.0 & 1.6 & $1.1\times10^{18}$ & 8.8 [14] & ... & ... & ... & Transient Z-source & 15,16\\
\exo & 54714.0 & 24 & $4.8\times10^{16}$ & 7.1 [17] & 552 [18] & 3.8 [19] & 94.6 & Eclipsing & 20,21,22,23\\
\igr & 55556.0 & 0.17 & $2.3\times10^{17}$ & 5.5 [24] & 11 [25] & 21.3 [25] & 73.6 & In Terzan 5 & 26,27,28,29 \\
 &  &  &  &  &  &  &  &  Transient Z-source & \\
 &  &  &  &  &  &  &  &  X-ray pulsar & \\
\maxisource & 56052.1 & 1.3 & $9.1\times10^{17}$ & 45 [30] & ... & 16.4/9.8 [31] & ... & Transient Z-source & 30\\
\source & 56166.0 & 0.15 & $9.6\times10^{16}$ & 5.5 [24] & ... & ... & 89.9 & In Terzan 5 & 29 \\
\hline
\end{tabular*}
\label{tab:sources}
\begin{tablenotes}
\item[]Note. -- The assumed end of the outburst is indicated by $t_0$, whereas $t_{\mathrm{ob}}$ and $\dot{M}_{\mathrm{ob}}$ are estimates of the outburst duration and average mass-accretion rate, respectively. Any available information on the distance ($D$), neutron star spin frequency ($\nu_{\mathrm{spin}}$), orbital period ($P_{\mathrm{orb}}$), and temperature measured before the outburst ($kT^{\infty}_{\mathrm{base}}$) are indicated with the appropriate references given in parenthesis. Type-I X-ray bursts have been detected from all sources except \maxisource. References: 1 = \citet{muno2000}, 2= \citet{smith1997}, 3 = \citet{wijnands2001}, 4 = \citet{wijnands2003}, 5 = \citet{cackett2006}, 6 = \citet{cackett2010}, 7 = \citet{muno2001}, 8 = \citet{wijnands01}, 9 = \citet{cominsky1984}, 10 = \citet{wijnands2002},  11 = \citet{wijnands2004}, 12 = \citet{cackett2008}, 13 = \citet{cackett2013_1659}, 14 = \citet{lin2009}, 15 = \citet{fridriksson2010}, 16 = \citet{fridriksson2011}, 17 = \citet{galloway06}, 18 = \citet{galloway2010}, 19 = \citet{parmar1986}, 20 = \citet{degenaar09_exo1}, 21 = \citet{degenaar2010_exo2}, 22 = \citet{degenaar2014_exo3}, 23 = \citet{diaztrigo2011}, 24 = \citet{ortolani2007}, 25 = \citet{papitto2010}, 26 = \citet{degenaar2011_terzan5_2}, 27 = \citet{degenaar2011_terzan5_3}, 28 = \citet{degenaar2013_terzan5}, 29 = this work, 30 = \citet{homan2014}, 31 = \citet{cornelisse2012}.
\end{tablenotes}
\end{threeparttable}
\end{table*}

\section{Introduction}~\label{sec:introduction}
Low-mass X-ray binaries (LMXBs) are excellent laboratories to gain insight into the structure and composition of neutron stars \citep[e.g.,][for recent reviews]{miller2013_NSreview,ozel2013_NSreview}. One promising avenue is to study how neutron stars are heated during transient accretion events and cool thereafter. This thermal response can give detailed information about the structure and composition of the crust, as well as the density and superfluid properties of the core \citep[e.g.,][]{rutledge2002,shternin07,brown08,page2012,turlione2013,wijnands2012,horowitz2015,medin2015}.

Transient LMXBs undergo occasional outbursts during which matter from the $\lesssim$$1~\Msun$ companion is rapidly accreted on to the neutron star. The accumulation of matter compresses the stellar crust and causes a series of nuclear reactions that include electron captures and density-driven fusion reactions \citep[e.g.,][]{haensel1990a,haensel2008,steiner2012}. The resulting energy release of $\simeq$2~MeV per accreted nucleon causes the crust to heat up \citep[][]{brown1998}, creating a temperature profile that depends on the location and magnitude of the heat sources, as well as the duration and brightness of the accretion episode \citep[e.g.,][]{brown08,page2013}. Typical outbursts reach a luminosity of $L_{\mathrm{X}} \simeq 10^{36-38}~\lum$ (0.5--10 keV) and last for weeks to months (e.g., such as seen in Aql X-1, \sax, and \exoter), although a handful of neutron stars continue to accrete for years to decades at this level (such as e.g., \mxb, \exo, and \hete).

Outbursts are usually separated by long (years to decades) periods of quiescence with a much lower X-ray luminosity of $L_{\mathrm{X}} \simeq 10^{31-34}~\lum$, suggesting that the mass-accretion rate on to the neutron star is strongly reduced. The heated crust can then cool by thermally conducting the energy gained during the outburst into the neutron star core and towards the surface, where it is radiated as neutrinos and (X-ray) photons, respectively. This thermal relaxation depends on the structure, thickness, and composition of the crust since these set the thermal transport properties \citep[e.g.,][]{lattimer1994,rutledge2002,brown08,medin2015,horowitz2015}, as well as the temperature of the stellar core, which relates to its density and superfluid properties \citep[e.g.,][]{shternin07,page2013}.

The X-ray emission observed during outbursts is dominated by the accretion flow and hence heating of neutron stars cannot be directly observed. However, thermal X-ray emission from the incandescent neutron star can often be detected in quiescence. This offers the opportunity to probe its thermal response to the accretion episode and to monitor the subsequent cooling. 

In the past decade, crust cooling has successfully been monitored for six transiently accreting neutron stars (Table~\ref{tab:sources}). In these LMXBs, the temperature of the neutron star has been seen to decrease systematically for several years following an accretion outburst. Comparing these observations with crust-cooling simulations has provided valuable new insight into the properties of neutron star crusts. For example, it has been established that heat is conducted rapidly and hence that the atomic nuclei in the crust must have a highly ordered structure \citep[][]{shternin07,brown08,page2013,turlione2013}. Furthermore, crust-cooling studies have the potential to place strong constraints on the composition of the outer crustal layers \citep[e.g.,][]{wijnands2012,medin2015}, to test the occurrence of non-spherical nuclear shapes (`nuclear pasta') in the deep inner crust \citep[e.g.,][]{horowitz2015}, and even to probe the superfluid properties of the ultradense stellar core \citep[e.g.,][]{page2013}.

Despite these successes, several open questions remain. One of the current puzzles is that modelling of the crust-cooling curve points, in some sources, to the presence of an extra source of energy (possibly as high as a few MeV~nucleon$^{-1}$) that is not accounted for by current nuclear heating models, but must be located at a shallow depth in the crust \citep[e.g., in \mxb\ and \igr;][]{brown08,degenaar2011_terzan5_3}. Understanding the origin of this shallow heating is of particular interest because it may also explain the occurrence of highly energetic thermonuclear bursts called `superbursts', which have ignition properties that require a considerably higher crust temperature than can be achieved with current nuclear heating models \citep[e.g., in 4U 1608--52 and \exoter;][]{cumming06,keek2008_1608,altamirano2012}. Furthermore, the transition between stable and unstable thermonuclear burning on the surface of neutron stars, which gives rises to mHz quasi-periodic oscillations, is observed to occur at an accretion rate that is about an order of magnitude lower than predicted by theory \citep[e.g., in Aql X-1, 4U 1608--52, 4U 1636--53, and \igr;][]{revnivtsev2001,altamirano2008,keek2009,linares2012_ter5_2}. This discrepancy could also be reconciled if additional crustal heating occurs. 

Possible explanations for the origin of extra heating include additional electron captures at shallow crustal layers \citep[e.g.,][]{estrade2011}, and previously unaccounted nuclear fusion reactions occurring deep in the crust \citep[e.g.,][]{horowitz2008}. However, electron captures may be too weak and nuclear fusion may occur at too high density to be reconciled with observational constraints on the extra heat release \citep[][]{degenaar2013_xtej1709}. An alternative mechanism that may be consistent with observations is a convective heat flux driven by the separation of light and heavy nuclei in the outer crustal layers \citep[][]{medin2011,medin2015,degenaar2013_xtej1709,degenaar2014_exo3}. Currently it is unclear whether extra crustal energy release occurs for all neutron stars, and if the depth and magnitude are always the same. Performing crust-cooling studies for additional sources, in particular probing a range of source parameters (e.g., outburst duration and brightness, orbital and spin period) can potentially shed light on this unresolved problem.

Crust-cooling studies initially focused on neutron stars exhibiting long ($>$1~yr) accretion outbursts to ensure that significant heating occurred and hence optimizing the chances of detecting the subsequent cooling. However, it has been demonstrated that significant heating can also occur during shorter outbursts \citep[in \igr;][]{degenaar2011_terzan5_2,degenaar2011_terzan5_3,degenaar2013_terzan5}. Following neutron stars after weeks to months long accretion outbursts may be particularly suited to investigate the presence of shallow heat, since the subsequent cooling should be faster \citep[e.g.,][]{brown08} and detailed modelling of the complete cooling curve can therefore start sooner. Moreover, short outbursts are much more common than prolonged ones and hence may offer better opportunities to increase the number of observed crust-cooling curves.

 \begin{figure}
 \begin{center}
	\includegraphics[width=8.0cm]{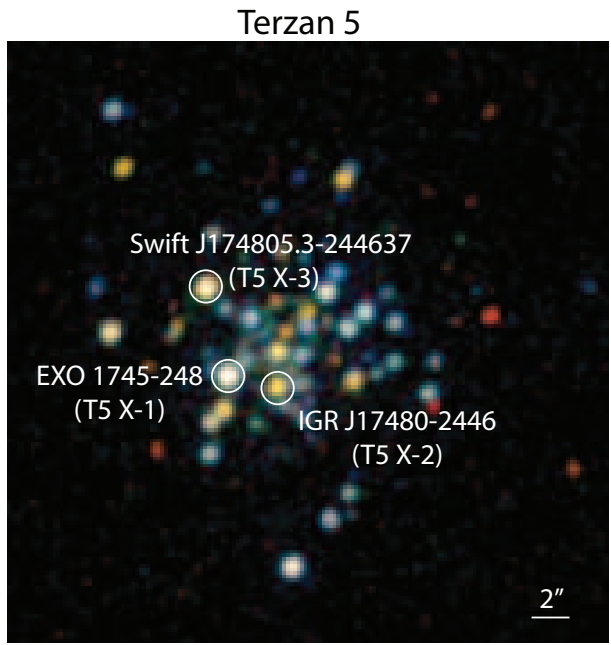}
    \end{center}
    \caption[]{Accumulated three-colour \chan/ACIS image of the inner $\simeq$1.1 arcmin $\times$1.1 arcmin of the globular cluster Terzan 5 (0.3--8~keV). Red colour corresponds to 0.3--1.5 keV, green to 1.5--2.5 keV, and blue to 2.5--8 keV. This image was constructed using 16 observations in which all bright X-ray transients were quiescent, amounting to 620 ks of total exposure time. The positions of the three known transient neutron star LMXBs are indicated by circular regions with 1.5 arcsec radii.
        }
 \label{fig:image}
\end{figure}

\subsection{\source\ in Terzan 5}
Terzan 5 is a strongly absorbed \citep[$N_{\mathrm{H}} \simeq 10^{22}~\nh$;][]{heinke2006_terzan5,bahramian2014} Milky Way globular cluster that is located at an estimated distance of $\simeq$4.6--8.7~kpc \citep[e.g.,][]{cohn2002,ortolani2007}. It harbours a large number of faint X-ray point sources, many of which are candidate quiescent neutron star LMXBs \citep[][see also Fig.~\ref{fig:image}]{heinke2006_terzan5}. Indeed, two confirmed transient neutron star LMXBs were previously identified in the cluster: \exoter\ and the 11-Hz X-ray pulsar \igr. The former was active at least in 2000, 2011, and 2015 \citep[e.g.,][]{heinke2003,altamirano2012,barret2012,serino2012,altamirano2015_ter5,tremou2015}, whereas the latter was responsible for the 2010 outburst of Terzan 5 \citep[e.g.,][]{papitto2010,miller2011,motta2011}. X-ray activity was also observed from the cluster in 1980, 1984, 1990, 1991, and 2002 \citep[see][for a historic overview up until 2012]{degenaar2012_1745}, but due to the high source density and lack of sub-arcsecond spatial resolution observations, it cannot be pinpointed with certainty which object(s) caused these outbursts (see Fig.~\ref{fig:image}). In 2012, a third transient neutron star LMXB was discovered in Terzan 5: \source\ \citep[also known as T5 X-3;][]{bahramian2014}. The cluster core with the locations of the three transient LMXBs is shown in Fig.~\ref{fig:image}.

The 2012 discovery outburst of \source\ started in early July and the source remained active for $\simeq$7--8~weeks. During this time it was detected at a mean 0.5--10 keV luminosity of $L_{\mathrm{X}}\simeq 9 \times 10^{36}~\dist~\lum$, and it displayed a thermonuclear X-ray burst \citep[][]{bahramian2014}. Analysis of archival \chan\ data revealed that the pre-outburst quiescent X-ray spectrum was dominated by thermal emission from the neutron star surface, but with an $\simeq$20--30 per cent contribution from a non-thermal emission component. The neutron star temperature remained constant between 2003 and 2012, but small variations in the quiescent non-thermal emission were observed \citep[][]{bahramian2014}.\footnote{No outburst was detected from \source\ in the epoch 2003--2012, suggesting that all these observations sampled the same quiescent episode. \citet{bahramian2014} did propose that because of the relatively high temperature of the neutron star, \source\ may have been responsible for one or more of the historic outbursts of Terzan 5 for which no accurate position information was available.} Such a hard (variable) emission tail is often detected for quiescent neutron star LMXBs and is typically modelled by a simple power law with an index of $\Gamma \simeq 1-2$. It is thought to arise from a different emission process, possibly related to the presence of a residual accretion stream or the magnetic field of the neutron star \citep[e.g.,][]{campana1998,jonker2004,wijnands05_amxps,cackett2010_cenx4,degenaar2012_amxp,bernardini2013,chakrabarty2014_cenx4}.

When the 2012 outburst of \source\ ceased and the source transitioned to quiescence, we took the opportunity to search for crustal cooling in this neutron star.

%%%%%%%%%%%%%%%%%
% OBSERVATIONS
%%%%%%%%%%%%%%%%%

\section{Data analysis and results}
\subsection{Observations and analysis procedures}
Terzan 5 has been observed five times since the end of the 2012 outburst of \source\ as part of a \chan\ Target of Opportunity programme to monitor the thermal evolution of the neutron star. Two additional \chan\ observations were performed to study the 11-Hz X-ray pulsar in quiescence \citep[][]{degenaar2013_terzan5}. Together these seven observations cover $\simeq$2~yr after the discovery outburst of \source\ (Table~\ref{tab:obs}). For all observations the ACIS-S3 chip was used with the `faint' timed data mode, either in full frame or in a 1/4 sub-array. None of the observations suffered from background flares. We reduced and analysed the data using the \textsc{ciao} tools version 4.4 and \textsc{caldb} version 4.5.2. 

Source events were obtained by using a circular region with a radius of 1.5 arcsec (see Fig.~\ref{fig:image}). A source-free circular region with a radius of 40 arcsec placed $\simeq$2 arcmin from the cluster core was used to extract background events. For each observation we extracted count rates using \textsc{dmextract} (Table~\ref{tab:obs}), whereas the spectra and response files were generated with \textsc{specextract}. After grouping the spectra with \textsc{grppha} into bins of $>$15 photons, we fitted the data in the 0.3--10 keV range using \textsc{xspec} \citep[version 12.8;][]{xspec}. Throughout this work we assume a distance of $D=5.5$~kpc for Terzan 5 \citep[][]{ortolani2007}, and quote uncertainties at the 1$\sigma$ level of confidence.

During the three exposures that were taken within $\simeq20$~d from one another in 2013 February, the count rates of \source\ were similar (Table~\ref{tab:obs}). To improve the statistics we combined these three spectra (and the weighted response files) using \textsc{combine$\_$spectra}. Similarly, the three exposures obtained in 2014 July had comparable count rates and were therefore combined (Table~\ref{tab:obs}). This resulted in post-outburst spectra for five different epochs.

To study any possible thermal response of the neutron star to the 2012 outburst, we also analyse the spectra of all pre-outburst observations in this work \citep[Table~\ref{tab:obs}, see also][]{bahramian2014}. These data were reduced in the same way as detailed above. The two exposures of 2011 September (Table~\ref{tab:obs}) were combined to create a single spectrum using \textsc{combine$\_$spectra}. This left us with seven pre-outburst spectra.

\begin{table}
\caption{\chan/ACIS-S observations of Terzan 5.}
\begin{threeparttable}
\begin{tabular*}{0.47\textwidth}{@{\extracolsep{\fill}}l c c c}
\hline 
ObsID & Date & Exposure Time  & Count rate \\
& & (ks) &  ($10^{-3}~\cnts$) \\
\hline
\multicolumn{4}{c}{Pre-outburst observations} \\	% rates from Bahramian 2014
3798 & 2003 Jul 13/14 & 39.5 & $10.0\pm 1.0$  \\ 	%9.91E-3
10059 & 2009 Jul 14/15 & 36.4 & $7.3\pm 0.5$  \\  %7.42E-3
13225 & 2011 Feb 17 & 29.7 & $5.0\pm 0.4$  \\	%5.56E-3
13252 & 2011 Apr 29/30 & 39.5 & $6.9\pm 0.4$  \\	%6.78E-3 
13705$^*$ & 2011 Sep 5  & 13.9 & $5.5\pm 0.6$ \\		%5.69E-3
14339$^*$ & 2011 Sep 8  & 34.1 & $6.0\pm 0.5$ \\		%6.08E-3
12454 & 2011 Nov 3 & 9.8 & $7.3\pm 0.8$ \\
13706 & 2012 May 13 & $46.5$ & $7.7\pm 0.4$ \\	%7.58E-3
%\hline
\multicolumn{4}{c}{Post-outburst observations} \\
14475 & 2012 Sep 17/18 & $30.5$ & $10.0\pm 0.6$ \\	
14476 & 2012 Oct 28 & $28.6$ & $9.4\pm 0.6$ \\	
14477$^*$ & 2013 Feb 5  & $28.6$ & $7.4\pm 0.5$ \\	
14625$^*$ & 2013 Feb 22  & $49.2$ & $7.1\pm 0.4$ \\	
15615$^*$ & 2013 Feb 24  & $84.2$ & $6.4\pm 0.3$ \\
14478 & 2013 Jul 16/17 & $28.6$ & $7.3\pm 0.5$ \\
14479$^*$ & 2014 Jul 15 & $28.6$ & $5.7\pm 0.5$ \\	
16638$^*$ & 2014 Jul 17 & $71.6$ & $5.6\pm 0.3$ \\	
15750$^*$ & 2014 Jul 20 & $23.0$ & $6.1\pm 0.5$ \\	
\hline
\end{tabular*}
\label{tab:obs}
\begin{tablenotes}
\item[]Note. -- The count rates for \source\ are given for the 0.3--10 keV energy range. Quoted uncertainties are at the 1$\sigma$ level of confidence. Spectra extracted from observations spaced close in time are marked by an asterisk and were combined to improve the statistics.
\end{tablenotes}
\end{threeparttable}
\end{table}

\begin{table*}
\caption{Spectral analysis results for \source\ in quiescence.\label{tab:spec}}
\begin{threeparttable}
\begin{tabular*}{0.96\textwidth}{@{\extracolsep{\fill}}cccccccccc}
\hline
Epoch & MJD & $kT^{\infty}$  & $F_{\mathrm{X}}$ & $F_{\mathrm{X,pow}}$  & $F_{\mathrm{th,bol}}$ & $f_{\mathrm{pow}}$ & $L_{\mathrm{X}}$ & $L_{\mathrm{th,bol}}$ \\
& & (eV) & \multicolumn{3}{c}{($10^{-13}~\flux)$} & & \multicolumn{2}{c}{($10^{33}~\lum)$} \\
\hline
\multicolumn{9}{c}{Pre-outburst} \\	
2003 Jul & 52833.5 & $89.7 \pm 1.7$ & $5.0 \pm 0.6$ & $1.53 \pm 0.17$ & $3.7 \pm 0.8$ & $0.30\pm0.16$ & $1.8 \pm 0.2$ & $1.4 \pm 0.3$\\
2009 Jul & 55027.5 & $89.7 \pm 1.7$ & $3.7 \pm 0.5$ & $1.02 \pm 0.14$ & $3.7\pm 0.8$ & $0.28\pm0.19$ & $1.3 \pm 0.2$ & $1.4 \pm 0.3$\\
2011 Feb & 55609 & $89.7 \pm 1.7$ & $3.2 \pm 0.4$ & $0.58 \pm 0.12$  & $3.7 \pm 0.8$& $0.18\pm0.25$ & $1.2 \pm 0.2$ & $1.4 \pm 0.3$ \\
2011 Apr & 55680.5 & $89.7 \pm 1.7$ & $3.7 \pm 0.4$ & $0.89 \pm 0.13$  & $3.7 \pm 0.8$& $0.24\pm0.19$ & $1.4 \pm 0.1$ & $1.4 \pm 0.3$ \\
2011 Sep & 55810.5 & $89.7 \pm 1.7$ & $2.9 \pm 0.4$ & $0.61 \pm 0.11$  & $3.7 \pm 0.8$& $0.21\pm0.22$ & $1.0 \pm 0.1$ & $1.4 \pm 0.3$\\
2011 Nov & 55868 & $89.7 \pm 1.7$ & $3.8 \pm 1.0$ & $0.78 \pm 0.30$  & $3.7 \pm 0.8$& $0.21\pm0.48$ & $1.4 \pm 0.4$ & $1.4 \pm 0.3$\\
2012 May & 56060 & $89.7 \pm 1.7$ & $4.1 \pm 0.5$  & $1.22 \pm 0.14$ & $3.7 \pm 0.8$& $0.30\pm0.16$ & $1.5 \pm 0.2$ & $1.4 \pm 0.3$ \\
%\hline
\multicolumn{9}{c}{Post-outburst} \\	
2012 Sep  & 56187.5 & $98.2 \pm 2.2$ & $4.9 \pm 0.6$ & $0.86 \pm 0.08$  & $5.1 \pm 0.7$& $0.18\pm0.23$ & $1.8 \pm 0.2$ & $1.8 \pm 0.2$ \\
2012 Oct  & 56228 & $93.9 \pm 2.6$ & $4.5 \pm 0.6$ & $1.18 \pm 0.20$  & $4.3 \pm 0.7$& $0.26\pm0.21$ & $1.6 \pm 0.2$ & $1.5 \pm 0.3$ \\
2013 Feb & 56340 & $90.0 \pm 1.6$ & $3.5 \pm 0.3$ & $0.72 \pm 0.08$  & $3.6 \pm 0.4$& $0.21\pm0.14$ & $1.3 \pm 0.1$ & $1.3 \pm 0.2$ \\
2013 Jul & 56489.5 & $90.1 \pm 2.7$ & $4.1 \pm 0.6$ & $1.18 \pm 0.19$  & $3.8 \pm 0.7$& $0.29\pm0.21$ & $1.5 \pm 0.2$ & $1.4 \pm 0.2$ \\
2014 Jul & 56855.5 & $88.0 \pm 1.9$ & $3.3 \pm 0.3$ & $0.78 \pm 0.09$  & $3.3 \pm 0.4$& $0.24\pm0.15$ & $1.1 \pm 0.1$ & $1.2 \pm 0.2$ \\
\hline
\end{tabular*}
\begin{tablenotes}
\item[]Note. -- $F_{\mathrm{X}}$ and $F_{\mathrm{X,pow}}$ represent the total unabsorbed model flux and the power-law flux in the 0.5--10 keV band, respectively. The parameter $f_{\mathrm{pow}}$ represents the fractional contribution of the power-law component to the total unabsorbed 0.5--10 keV flux. $F_{\mathrm{th,bol}}$ is the flux of the thermal component in the 0.01--100 keV range. $L_{\mathrm{X}}$ and $L_{\mathrm{th,bol}}$ are the total 0.5--10 keV and thermal 0.01--100 keV luminosities for $D=5.5$~kpc, respectively. The following parameters were kept fixed during the fits: $M=1.4~\Msun$, $R=10$~km, $D=5.5$~kpc, and $N_{\mathrm{nsatmos}}=1$. The simultaneous fit resulted in $N_{\mathrm{H}}=(2.63\pm0.10) \times 10^{22}~\nh$, $\Gamma=1.7\pm0.2$, and $\chi_{\nu}^2=1.03$ (for 162 dof). 
%Quoted errors are 1$\sigma$.
\end{tablenotes}
\end{threeparttable}
\end{table*}

%%%%%%%%%%%%%%%%%
% SPECTRA
%%%%%%%%%%%%%%%%%

\subsection{X-ray spectral analysis}
Following \citet{bahramian2014}, we used a combination of a neutron star atmosphere model \citep[\textsc{nsatmos};][]{heinke2006}, and a power-law (\textsc{pegpwrlw}, where we set the normalization to represent the unabsorbed 0.5--10 keV flux) to fit all spectra. For the \textsc{nsatmos} component we fix the neutron star mass ($M=1.4~\Msun$) and radius ($R=10$~km), the distance ($D=5.5$~kpc), and the normalization ($N_{\mathrm{nsatmos}}=1$, i.e., the emission radius was assumed to correspond to the entire neutron star surface). The data did not provide sufficient statistics to constrain variations of the power-law index when it was allowed to vary between the different observations. We therefore assumed this parameter to be constant over the entire data set and only allowed the power-law normalization to vary. To account for interstellar absorption we included the \textsc{tbabs} model, adopting the \textsc{vern} cross-sections and \textsc{wilm} abundances \citep[][]{verner1996,wilms2000}. The hydrogen column density was assumed to remain constant at all epochs (i.e., this parameter was tied between all data sets), which seems reasonable since the source did not display any eclipses or dips and is therefore likely not viewed at very high inclination \citep[see][]{miller2009}.

\begin{figure}
 \begin{center}
	\includegraphics[width=8.5cm]{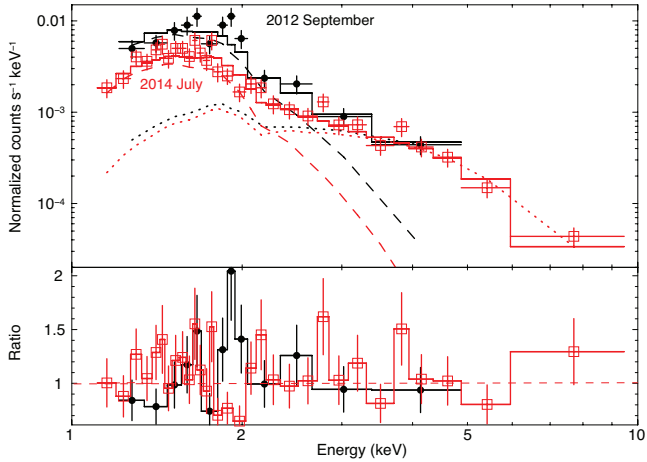}
    \end{center}
    \caption[]{Comparison between the post-outburst X-ray spectra of 2012 September (black, filled circles) and 2014 July (red, open squares) for \source. The solid lines represent fits using a model consisting of a neutron star atmosphere (dashed curves) and power law (dotted curves). The bottom plot shows the data to model ratio of both spectra.
        }
 \label{fig:spec}
\end{figure}

We first addressed whether there was evidence for variability among the five post-outburst spectra. Assuming that both the thermal and the non-thermal model components were the same at all epochs resulted in a poor fit ($\chi_{\nu}^2=1.51$ for 105 dof; with a $p$-value of $P_{\chi}=5.9\times10^{-4}$). Allowing the normalization of the power law to vary improved the fit ($\chi_{\nu}^2=1.20$ for 101 dof, $P_{\chi}=0.08$). However, a more significant improvement was achieved by allowing the temperature to vary (with the power-law normalization tied; $\chi_{\nu}^2=1.10$ for 101 dof, $P_{\chi}=0.22$). Leaving the power-law normalization free in addition to the temperature, further improved the fit ($\chi_{\nu}^2=1.02$ for 97 dof, $P_{\chi}=0.41$), although an $F$-test suggests a probability of 0.02 that such an improvement occurs by chance. In Fig.~\ref{fig:spec} we show the X-ray spectra of the first (2012 September) and last (2014 July) post-outburst observations. This plot illustrates that the largest spectral changes occurred below $\simeq$3~keV, where the thermal emission from the neutron star dominates (dashed curves in Fig.~\ref{fig:spec}). We conclude that there is strong evidence for a changing neutron star temperature after the 2012 outburst, but there were no large variations in the non-thermal power-law emission.

\begin{figure}
 \begin{center}
	\includegraphics[width=8.3cm]{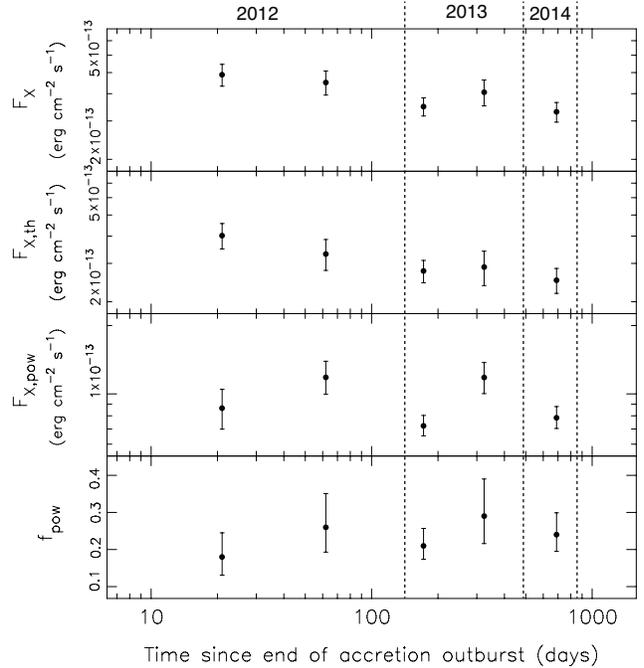}
    \end{center}
    \caption[]{Flux evolution in the 0.5--10 keV band after the 2012 outburst of \source. From top to bottom: the total unabsorbed flux ($F_{\mathrm{X}}$), thermal flux ($F_{\mathrm{X,th}}$), power-law flux ($F_{\mathrm{X,pow}}$), and power-law fraction ($f_{\mathrm{pow}}$). The vertical dotted lines and the numbering on top indicate the different years of the monitoring campaign. 
    %Error bars indicate 1-$\sigma$ confidence intervals. 
    }
 \label{fig:fluxes}
\end{figure}

We proceeded to fit the entire set of twelve (seven pre-outburst, five post-outburst) quiescent observations, each with their own response files, together in \textsc{xspec}. Since \citet{bahramian2014} found no evidence for changing thermal emission before the outburst, we only allowed the neutron star temperature to vary independently for the new data (i.e., it was assumed to be the same for all pre-outburst observations). We determined for each observation: the neutron star temperature (as observed at infinity $kT^{\infty}$),\footnote{The temperature seen by an observer at infinity relates to the fitted temperature, $kT$, as $kT^{\infty}= kT/(1+z)$, where $1+z = (1-R_{\mathrm{s}}/R)^{-1/2} = 1.31$ is the gravitational redshift for our choice of $M=1.4~\Msun$ and $R=10$~km. Here $R_{\mathrm{s}}=2GM/c^2$ is the Schwarzschild radius, $G$ the gravitational constant, and $c$ the speed of light.} the total unabsorbed model flux $F_{\mathrm{X}}$, the power-law flux $F_{\mathrm{X,pow}}$, and the relative contribution of the power-law flux $f_{\mathrm{pow}}$ (all 0.5--10 keV), as well as the thermal bolometric flux $F_{\mathrm{th,bol}}$ (0.01--100~keV). The fluxes and corresponding errors were calculated using the \textsc{cflux} command in \textsc{xspec}. The results of our spectral analysis are summarized in Table~\ref{tab:spec}, and the flux evolution is shown in Fig.~\ref{fig:fluxes}.

Our approach resulted in a good fit ($\chi_{\nu}^2=1.03$ for 162 dof, $P_{\chi}=0.37$). We obtained a hydrogen column density of $N_{\mathrm{H}} = (2.63\pm 0.10)\times 10^{22}~\nh$, and a power-law index of $\Gamma=1.7\pm 0.2$. Both are consistent with the pre-outburst analysis of \citet{bahramian2014}. All quiescent spectra of \source\ are dominated by thermal emission, which contributes $\simeq70-80$ per cent to the total unabsorbed 0.5--10 keV flux. The fractional contribution of the power-law component does not show large variations between the different observations (see Fig.~\ref{fig:fluxes} and Table~\ref{tab:spec}). In fact, this parameter can be adequately fitted with a constant function, yielding $f_{\mathrm{pow}}=0.24\pm0.01$ in 2012--2014.

For the thermal emission component we measure a pre-outburst temperature of $kT^{\infty}_{\mathrm{base}}= 89.7 \pm 1.7$~eV. During the first two new observations, obtained $\simeq$2~weeks and $\simeq$2~months after the 2012 outburst, the temperature was elevated to $kT^{\infty}= 98.2 \pm 2.2$~eV and $kT^{\infty}= 93.9 \pm 2.6$~eV, respectively. The subsequent three observations, obtained $\simeq$0.5--2~yr after the outburst, are all consistent with the pre-outburst temperature (Table~\ref{tab:spec}). Indeed, fixing the temperatures of the last three observations to those of the pre-outburst spectra provides an acceptable fit ($\chi_{\nu}^2=1.03$ for 165 dof, $P_{\chi}=0.37$). 

\begin{figure}
 \begin{center}
	\includegraphics[width=8.3cm]{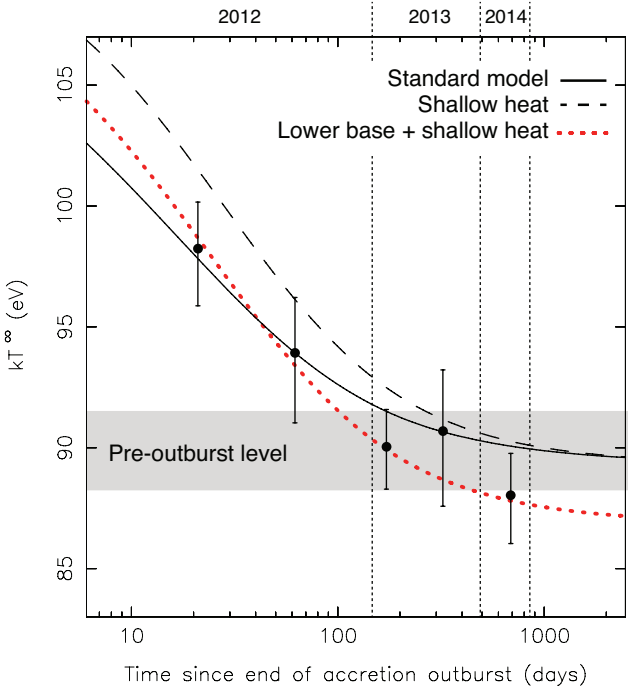}
    \end{center}
    \caption[]{Thermal evolution of \source\ compared to model calculations. The observations can be described with standard physics input (black solid curve), but a shallow heat source up to $1.4$~MeV~nucleon$^{-1}$ is allowed at an approximate 2$\sigma$ level (dashed curve). The dotted curve shows a model calculation with a lower base level, in which case a shallow heat source of  $0.85$~MeV~nucleon$^{-1}$ is preferred by the data (see Section~\ref{subsec:model} for details). The grey horizontal bar represents the temperature measured prior to the 2012 outburst. The vertical dotted lines and the numbering on top indicate the different years in the campaign. 
    %Error bars indicate 1-$\sigma$ confidence intervals. 
    }
 \label{fig:model}
\end{figure}

An $F$-test suggests that leaving the temperature of the second observation free yields a $\simeq1.5\sigma$ improvement compared to fixing it to the pre-outburst spectra. Leaving the temperature of the first observation free results in an $\simeq4\sigma$ improvement. We conclude that there is firm evidence for an enhanced temperature during the first post-outburst observation, but only a marginal indication that it was still enhanced during the second observation. Nevertheless, Fig.~\ref{fig:model} illustrates that there is a systematic decrease in temperature following the 2012 outburst. Such a temperature evolution is not seen in the pre-outburst data \citep[][]{bahramian2014}. In analogy with that seen in other sources (see Section~\ref{subsec:comparison}), this may point to crust cooling. It would imply that the neutron star crust was heated during the $\simeq8$~week accretion outburst but cooled rapidly (within $\simeq$100~d) afterwards. 

In Section~\ref{subsec:model}, we compare the observed temperature curve of \source\ with thermal evolution simulations to study the crust properties of this neutron star. In Section~\ref{subsec:curve}, we also report on empirical decay fits to the temperature curve to provide a comparison with other crust-cooling sources.

\begin{table*}
\caption{Model parameters for thermal evolution calculations of \source.\label{tab:model}}
\begin{threeparttable}
\begin{tabular*}{0.96\textwidth}{@{\extracolsep{\fill}}lccc}
\hline
Parameter (unit) / Model & Standard model & Shallow heat & Lower base + shallow heat \\
\hline
Impurity parameter, $Q_{\mathrm{imp}}$ & $1.0$ & $1.0$ & $1.0$   \\
Mass-accretion rate during outburst, $\dot{M}_{\mathrm{ob}}$ (g~s$^{-1}$) & $1.0\times10^{17}$ & $1.0\times10^{17}$ & $1.0\times10^{17}$  \\
Outburst duration, $t_{\mathrm{ob}}$ (yr) & $0.15$ & $0.15$ & $0.15$ \\
End of the outburst, $t_0$ (MJD) & 56166 & 56166 & 56166  \\
Atmosphere temperature during outburst, $T_{\mathrm{ob}}$ (K) & $2.8\times10^8$ & $2.8\times10^8$ & $2.8\times10^8$  \\
Core temperature, $T_{\mathrm{core}}$ (K) & $1.55\times10^8$ &  $1.55\times10^8$ &  $1.45\times10^8$ \\
Additional shallow heat, $Q_{\mathrm{shallow}}$ (MeV) & ... & 1.4 & 0.85 \\
Depth of shallow heat, $P/g$ (g~cm$^{-2}$) & ... & $1.0\times10^{13}$ & $1.0\times10^{13}$ \\
$\chi^2_{\nu}$ (dof) & 0.4 (4) & 1.5 (4) & 0.1 (4) \\
\hline
\end{tabular*}
\begin{tablenotes}
\item[]Note. -- The three different models are shown in Fig.~\ref{fig:model}. See Section~\ref{subsec:model} for details.
\end{tablenotes}
\end{threeparttable}
\end{table*}

%%%%%%%%%%%%%%%%%
% CRUST COOLING CURVE
%%%%%%%%%%%%%%%%%

\subsection{Crust-cooling simulations}\label{subsec:model}
To test the hypothesis that the crust of \source\ was significantly heated during its 2012 outburst and that our \chan\ observations track the subsequent cooling, we performed simulations with the neutron star thermal evolution code of \citet{brown08}, using all the physics input and modelling approach described in that work.\footnote{This code is available at: https://github.com/nworbde/dStar.} In brief: the crust composition was assumed to match the calculations of \citet{haensel1990a}, and the temperature in the atmosphere during outburst was set to $T_{\mathrm{ob}}=2.8\times10^{8}$~K. This is appropriate if H/He burning occurs, as indicated by the properties of a type-I X-ray burst observed from \source\ \citep[][]{bahramian2014}. 

The first and the last detection of the 2012 outburst (with \swift/XRT) were on July 6 and August 24, respectively. The source was not seen active yet on June 30, and on August 30 it had faded into the local background \citep[][]{bahramian2014}. The 2012 outburst thus had a duration of $\simeq$49--61~d ($t_{\mathrm{ob}} =$0.15$\pm0.02$~yr), and ended between 2012 August 24 and 30. During this time it was detected at an average 0.5--10 keV luminosity of $L_{\mathrm{X}} \simeq 9 \times 10^{36}~\dist~\lum$. Assuming a bolometric correction factor of 2 \citep[][]{zand07}, we estimate an average mass-accretion rate during the outburst of $\dot{M}_{\mathrm{ob}} = RL/GM \simeq 1.5\times10^{-9}~\mdot \simeq 9.6 \times 10^{16}~\mdotgs$ (for $M=1.4~\Msun$ and $R=10$~km). Fitting the 0.5--10 keV luminosity light curve of the outburst presented in \citet{bahramian2014} to a simple broken linear decay function suggests that the source faded into the XRT background $\simeq$36.5~d after the outburst peak, which was observed on MJD 56129.1. We therefore tentatively set the outburst end to $t_0 =~$MJD~56166. 

The temperature of the neutron star core was set to $T_{\mathrm{core}}=1.55\times10^{8}$~K to match the observed pre-outburst surface temperature ($kT^{\infty}=$89.7~eV). We assumed an impurity parameter $Q_{\mathrm{imp}}=1$, which corresponds to a highly conductive crust as appropriate for all other crust-cooling neutron stars \citep[e.g.,][]{brown08,page2013,degenaar2011_terzan5_3,degenaar2014_exo3,horowitz2015}. However, we note that as a result of the short outburst duration the cooling curves of \source\ are not sensitive to this parameter. This is because only the outer layers are heated, where the thermal conductivity is governed by electron-ion rather than electron-impurity scattering \citep[see also][]{degenaar2011_terzan5_3}.

Fig.~\ref{fig:model} shows that the observed temperature curve of \source\ can be reproduced using the standard physics input (solid curve), and does not require the inclusion of an additional source of shallow heat (Table~\ref{tab:model}). We briefly explored to what extent the presence of such shallow heating is allowed by the current data, using a grid search technique on $Q_\mathrm{{shallow}}$. For our standard model (i.e., no shallow heating), we found $\chi_{\nu}^2 = 0.4$ for 4 dof. Adding any source of shallow heat increases the $\chi^2$ value, hence gives a worse fit. By increasing the magnitude of the shallow heat until we obtain a fit statistic of $\Delta \chi^2 \simeq 4$ compared to the standard model, we can set an approximate 2$\sigma$ upper limit of $Q_\mathrm{{shallow}} \lesssim 1.4$~MeV~nucleon$^{-1}$ ($\chi_{\nu}^2 = 1.5$ for 4 dof; Table~\ref{tab:model}). Fig.~\ref{fig:model} illustrates that shallow heating most prominently increases the temperature at early times after the outburst (dashed curve). It is therefore that observations within $\simeq2$~months after the end of an outburst are instrumental to constrain shallow crust heating \citep[see also e.g.,][]{brown08,page2012}.

We probed the effect of uncertainties in the outburst properties of \source\ on our conclusions about the shallow  heating. Allowing for any realistic longer/shorter outburst duration ($t_{\mathrm{ob}}=0.13-0.17$~yr), a factor 2 higher/lower mass-accretion rate, or shifting the outburst end 3 d forward or backward ($t_0=$~MJD 56163--56169), changes the obtained limit on $Q_\mathrm{{shallow}}$ by $\pm0.3$~MeV~nucleon$^{-1}$ at most. However, a stronger effect is caused by changing the assumed base level. For instance, lowering the core temperature to $T_{\mathrm{core}}=1.45\times10^8$ (corresponding to an observed surface temperature of $\simeq$87~eV) actually prefers the addition of shallow heating, yielding a best fit for $Q_\mathrm{{shallow}}=0.85$~MeV~nucleon$^{-1}$ ($\chi_{\nu}^2 = 0.1$ for 4 dof; Table~\ref{tab:model}). This model is shown as the red dotted curve in Fig.~\ref{fig:model}. With this lower core temperature, we can set a 2$\sigma$ upper limit of $Q_\mathrm{{shallow}} \lesssim 2.4 $~MeV~nucleon$^{-1}$, i.e., $\simeq$1~MeV~nucleon$^{-1}$ higher than obtained for our standard model. Firmly establishing the base temperature during the current quiescent phase of \source\ is thus very important to accurately constrain the amount of shallow heating in this neutron star.

\subsection{Empirical decay fits to the temperature curve}\label{subsec:curve}
Observationally, the crust-cooling curves of \ks, \xte, \exo, and \igr\ can be described by a (broken) power-law decay \citep[e.g.,][]{cackett2010,fridriksson2011,degenaar2013_terzan5,degenaar2014_exo3}, whereas an exponential decay was found to be an adequate, or even better, description for \mxb\ and \maxisource\ \citep[][]{cackett2006,homan2014}. To compare the crust-cooling curve of \source\ with that of these other sources, we therefore fitted the temperature evolution to an exponential decay function of the form $y(t)=a~e^{-(t-t_0)/\tau}+b$, and a power-law decay of the form $y(t)=a~(t-t_0)^{-\alpha}+b$. Here, $a$ is a normalization constant, $b$ a constant offset that represents the quiescent base level (i.e., reflecting the core temperature of the neutron star), $\tau$ the e-folding time, $\alpha$ the decay index, and $t_0$ the start time of the cooling curve.  

\begin{figure}
 \begin{center}
	\includegraphics[width=8.3cm]{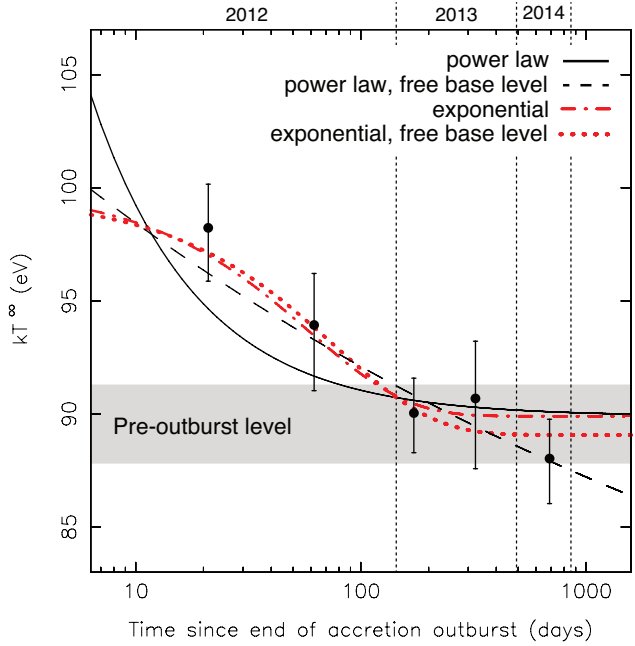}
    \end{center}
    \caption[]{Evolution of the neutron star temperature after the 2012 outburst of \source\ compared to empirical decay fits. The black solid (power-law) and red dash-dotted (exponential) fits assume that the source settles at its pre-outburst level (2003--2012; grey shaded area). The black dashed (power law) and red dotted (exponential) curves instead represent decay fits with the quiescent base level left as a free parameter. The vertical dotted lines indicate the different years in the campaign. 
    %Error bars indicate 1-$\sigma$ confidence intervals. 
        }
 \label{fig:temp}
\end{figure}

%To account for the effects of a different envelope composition after an outburst, and hence different observed base levels \citep[][]{brown2002,medin2015}, 
We performed fits both by fixing the base temperature at the value measured before the outburst ($kT^{\infty}_{\mathrm{base}}=89.7$~eV), and by leaving it free. The different decay fits are shown in Fig.~\ref{fig:temp} and the resulting fit parameters are given in Table~\ref{tab:lc}. For comparison with other sources we also quote the results for a power-law decay fit without a constant offset. All fits provide an adequate description of the present data; due to the limited number of data points and relatively large error bars it is not possible to prefer one statistically over the other. An exponential fit favours a base level that is close to the temperature measured before the outburst ($\simeq89.0$~eV; red dotted curve in Fig.~\ref{fig:temp}), whereas a power-law fit tends to a lower base level ($\simeq71.0$~eV; black dashed curve in Fig.~\ref{fig:temp}).

\begin{table}
\caption{Decay fits to the quiescent temperature curve of \source.\label{tab:lc}}
\begin{threeparttable}
\begin{tabular*}{0.45\textwidth}{@{\extracolsep{\fill}}lc}
\hline
Fit parameter (unit) & Value  \\
\hline
\multicolumn{2}{c}{{Exponential decay, base level fixed}}  \\
Normalization, $a$ (eV) & $10.1 \pm 3.3$   \\
Decay time, $\tau$ (days) & $59.3 \pm 41.9$  \\
Constant offset, $b$ (eV) & 89.7 fixed   \\
$\chi^2_{\nu}$ (dof) & 0.39 (3)   \\
$P_{\chi}$ & 0.76   \\
%\hline
\multicolumn{2}{c}{{Exponential decay, base level free}}  \\
Normalization, $a$ (eV)  & $10.6 \pm 2.9$   \\
Decay time, $\tau$ (days) & $77.7 \pm 49.1$  \\
Constant offset, $b$ (eV) & $89.0 \pm 1.0$  \\
$\chi^2_{\nu}$ (dof) & 0.20 (2)   \\
$P_{\chi}$ & 0.75   \\
%\hline
\multicolumn{2}{c}{{Power-law decay, base level fixed} } \\
Normalization, $a$ (eV) & $75.9 \pm 10.2$   \\
Decay index, $\alpha$ & $0.91 \pm 0.08$  \\
Constant offset, $b$ (eV) & 89.7 fixed   \\
$\chi^2_{\nu}$ (dof) & 0.70 (3)   \\
$P_{\chi}$ & 0.55   \\
%\hline
\multicolumn{2}{c}{{Power-law decay, base level free}}  \\
Normalization, $a$ (eV) & $35.8 \pm 4.9$  \\
Decay index, $\alpha$ & $0.11 \pm 0.03$ \\
Constant offset, $b$ (eV) & $70.9 \pm 22.1$   \\
$\chi^2_{\nu}$ (dof) & 0.16 (2)   \\
$P_{\chi}$ & 0.78   \\
%\hline
\multicolumn{2}{c}{{Power-law decay, no constant offset}}  \\
Normalization, $a$ (eV) & $104.1 \pm 3.7$  \\
Decay index, $\alpha$ & $0.03 \pm 0.01$  \\
Constant offset, $b$ (eV) & $0$ fix   \\
$\chi^2_{\nu}$ (dof) & 0.20 (3)   \\
$P_{\chi}$ & 0.89   \\
\hline
\end{tabular*}
\begin{tablenotes}
\item[]Note. -- The start of the cooling curve was set to  $t_0=$~MJD 56166. 
\end{tablenotes}
\end{threeparttable}
\end{table}

\subsection{A comparison with \igr}\label{subsec:comparison}
Crust cooling has been detected for six different sources before (Table~\ref{tab:sources}). Five of these exhibited a long outburst ($>$1~yr), providing the conditions for the crust to be significantly heated so that the subsequent cooling is observable. However, there is also strong evidence for crust cooling in the neutron star \igr\ following an $\simeq$2.5~month accretion outburst \citep[][]{degenaar2011_terzan5_2,degenaar2011_terzan5_3,degenaar2013_terzan5}. Given the similarly short outburst duration, it is therefore interesting to compare \source\ to this source in particular. 

\begin{figure}
 \begin{center}
	\includegraphics[width=8.3cm]{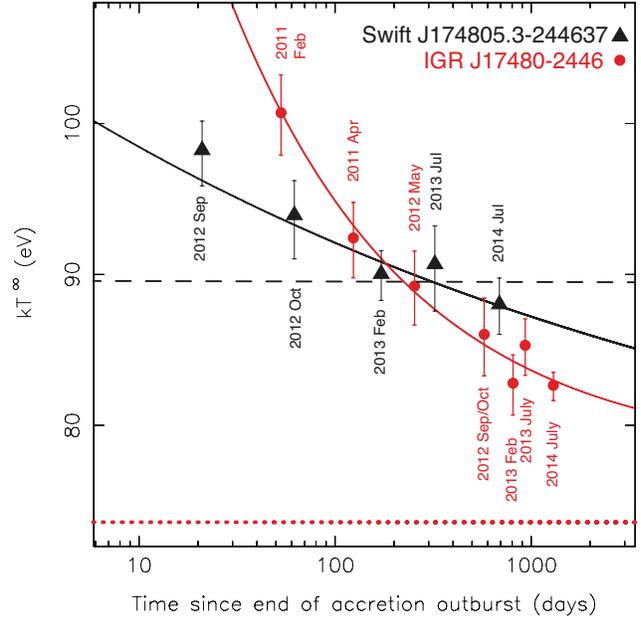}
    \end{center}
    \caption[]{ A comparison between the two short-duration transients for which crust cooling is observed (both located in the globular cluster Terzan 5). Black triangles are for \source, red filled circles are for \igr. Their pre-outburst base levels are indicated by the black dashed and red dotted horizontal lines, respectively. The solid curves indicate power-law decay fits with a free base level. 
    %Error bars indicate 1-$\sigma$ confidence intervals. 
    }
 \label{fig:compare}
\end{figure}

\igr\ is also located in the globular cluster Terzan 5 (see Fig.~\ref{fig:image}) and therefore has similar \chan\ coverage as \source. This 11-Hz X-ray pulsar exhibited an outburst in 2010 that lasted for $t_{\mathrm{ob}}\simeq77$~d \citep[][]{degenaar2011_terzan5_2}, with an estimated average accretion rate of $\dot{M} \simeq 2.3\times10^{17}$~g~s$^{-1}$ \citep[][]{degenaar2011_terzan5}. During the first three years in quiescence (2011--2013), the source showed a gradual decrease in temperature following a power-law decay with an index of $-\alpha = 0.47\pm0.05$, and a constant offset of $b=77.3\pm1.0$~eV, which is slightly higher than the measured pre-outburst temperature of $kT^{\infty}=73.6\pm1.6$~eV \citep[][see also Fig.~\ref{fig:compare}]{degenaar2013_terzan5}. Here, we update the cooling curve of \igr\ using the two new observations of 2013 July and 2014 July, which were obtained after the previous report by \citet{degenaar2013_terzan5}.

For the new data we followed the same analysis steps as for previous observations of \igr. We fitted the spectral data to an \textsc{nsatmos} model with $M=1.4~\Msun$, $R=10$~km, $D=5.5$~kpc, and $N_{\mathrm{H}} = 1.98 \times 10^{22}~\nh$ \citep[for details, see][]{degenaar2013_terzan5}. This resulted in neutron star temperatures of $kT^{\infty}=85.4 \pm 1.9$~eV (2013 July) and $kT^{\infty}=82.7 \pm 1.0$~eV (2014 July). The spectral fits do not require the inclusion of a hard emission component, suggesting a contribution of $\lesssim$5\% to the unabsorbed 0.5--10 keV flux. The updated cooling curve of \igr\ is shown in Fig.~\ref{fig:compare} (red data points). The two new measurements yielded a temperature similar to that of the preceding observation (2013 February). This is consistent with the slow temperature evolution expected at this stage. Indeed, fitting the updated cooling curve to a power-law decay gives $-\alpha = 0.45\pm 0.04$, and a constant offset of $b=77.3\pm 0.9$~eV (shown as the solid red curve in Fig.~\ref{fig:compare}). These decay parameters are essentially the same as obtained previously for this source \citep[quoted above;][]{degenaar2013_terzan5}, suggesting that in 2013--2014 the source remained on the cooling track seen in 2011--2013. 

To provide a direct comparison, we also show the post-outburst temperature evolution of \source\ in Fig.~\ref{fig:compare} (black data points and power-law decay curve). This illustrates that shortly after the outburst \source\ was less hot than \igr, which can plausibly be explained by its shorter outburst duration ($t_{\mathrm{ob}}\simeq 2$ versus $\simeq$2.5~months) and lower mass-accretion rate ($\dot{M}_{\mathrm{ob}} \simeq 1\times10^{17}$ versus $\simeq 2\times10^{17}$~g~s$^{-1}$). Moreover, the temperature curve of \source\ is flatter, which could be the result of its higher base level compared to \igr\ ($kT^{\infty}_{\mathrm{base}} \simeq 90$ versus $\simeq74$~eV; dashed and dotted horizontal lines in Fig.~\ref{fig:compare}). The fact that both sources show a continuous decay and that the temperature evolution of \igr\ can be well described by crust-cooling models \citep[][]{degenaar2013_terzan5}, lends support to the hypothesis that we also observe crust cooling in \source.

Since \source\ and \igr\ are both in the core of Terzan 5 and hence their distances should be the same, we can explore how their heat input during outburst and the output during the cooling phase compare (with the caveat that neutron star parameters also play a role in both the heating and the subsequent cooling). To this end we integrated the bolometric thermal flux in quiescence from $t=0$ till $t=700$~d since the outburst end (which corresponds to the time of the last available data point for \source), after subtracting the constant level due to the core temperature. We then obtain a ratio of the fluence in the cooling curve of \igr\ compared to \source\ of $f_{\mathrm{cool}}\simeq 5$. This is comparable to the ratio of the fluences of the outbursts of the two sources ($f_{\mathrm{ob}}\simeq 3$), as would be expected within the crustal cooling paradigm.

%%%%%%%%%%%%%%%%%
% DISCUSSION
%%%%%%%%%%%%%%%%%

\section{Discussion}\label{sec:discuss}
We reported on \chan\ monitoring of the transient neutron star LMXB \source\ in the globular cluster Terzan 5, following its $\simeq$8-week long 2012 discovery outburst. Observations were obtained in five different epochs in 2012--2014 and compared to seven X-ray spectra obtained prior to the accretion outburst in 2003--2012. All quiescent spectra are dominated by a soft, thermal component, while a non-thermal emission tail contributes $\simeq$20--30 per cent to the 0.5--10 keV unabsorbed flux. Analysis of the pre-outburst spectra revealed small variations in this hard power-law emission \citep[][]{bahramian2014}, but we detect no strong variability in this spectral component for the post-outburst data. 

Notably, the fractional contribution of the power-law spectral component to the total 0.5--10 keV flux was consistent with being constant in 2012--2014, despite an observed $\simeq$35 per cent variation in neutron star temperature (see below). This suggests that the quiescent thermal and power-law components are physically linked in \source. Similar results were obtained, for instance, for the neutron star LMXBs Cen X-4 \citep[][]{cackett2010_cenx4,bernardini2013} and \maxisource\ \citep[][]{homan2014}, whereas in other sources such as \xte\ \citep[][]{fridriksson2011} and \exo\ \citep[][]{degenaar2010_exo2} frequent quiescent monitoring has not revealed a clear connection between the two spectral components. A tight link between the quiescent thermal and non-thermal emission could suggest that (at least in some sources) both arise from residual accretion, although this would require fine-tuning  accretion models \citep[][]{cackett2010_cenx4,chakrabarty2014_cenx4}. Alternatively, the power-law tail may be an integral part of the neutron star spectrum that arises because of the limitations of current atmosphere models \citep[][]{homan2014}.

Analysis of the seven pre-outburst observations indicated that the temperature of the neutron star in \source\ remained constant \citep[][]{bahramian2014}. However, we find that $\simeq$2~weeks after the end of the 2012 outburst the neutron star temperature was significantly elevated compared to that measured before it became active. The lack of strong variability in the non-thermal spectral component implies that there is no obvious indication that this elevated temperature was due to continued low-level accretion. We instead propose that the crust of the neutron star in \source\ was heated during its 2012 accretion outburst and that our \chan\ observations sampled the subsequent cooling in quiescence.

\subsection{Crust cooling in \source}\label{subsec:crustcool}
If the crust of the neutron star was indeed significantly heated during the recent outburst, then our analysis suggests that it cooled rapidly, within $\simeq$100~d of entering quiescence. Fits to the temperature curve indicate an exponential decay time-scale of $\simeq$60--80~d, which is considerably shorter than the $>$400~d cooling time-scale of \ks\ and \mxb\ \citep[both these sources continue to cool many years after their outburst ended; e.g.,][]{cackett2008,cackett2010}. However, considering the typically large errors on the characteristic decay times (see e.g., Table~\ref{tab:lc}), the cooling time of \source\ is not significantly different from the $\simeq$150--200~d decays observed for \exo, \igr, and \maxisource\ \citep[][]{degenaar2013_terzan5,degenaar2014_exo3,homan2014}. On the other hand, the data of \source\ may also be consistent with continued cooling following a power-law decay, which would suggest a post-outburst base level $\simeq$20 per cent lower than observed prior to the 2012 accretion phase. This could potentially be explained as a different amount of H/He left on the surface of the neutron star after its last outburst \citep[][see also Section~\ref{subsec:extraheat}]{brown2002,medin2015}.

A direct comparison with crust cooling observed after a 10-week accretion outburst of \igr\ suggests that the more rapid decay and flatter temperature curve of \source\ may naturally be explained by its higher base temperature. As a result, the temperature profile in the crust at the end of the outburst is much less steep than for a longer outburst and/or a lower core temperature, so that the thermal relaxation is much quicker \citep[see e.g.,][]{brown08,page2012}. This may also be the reason that there is no clear evidence for (long-term) crust cooling in Aql X-1 \citep[][]{brown1998,rutledge2002_aqlX1}. Its outburst properties are quite similar to \source\ \citep[on average $t_{\mathrm{ob}}\simeq70$~d, $L_{\mathrm{X}}\simeq3\times10^{36}~\lum$;][]{campana2013}, but its quiescent temperature is considerably higher, $kT^{\infty}\simeq$110~eV \citep[e.g.,][]{cackett2011_aqlx1}. Neutron stars with low pre-outburst quiescent temperatures may therefore be the best targets for future studies, even if their outburst is short. 

It is of note that rapid temperature decays were seen at the end of the (short) outbursts of e.g., \rx\ \citep[][]{degenaar2013_xtej1709}, \igrmontse\ \citep[][]{armas2013_2},\footnote{\igrmontse\ is a very faint LMXB, not to be confused with the crust-cooling source \igr\ in Terzan 5.} and Aql X-1 \citep[][]{campana2014}. The characteristic time-scale in these sources is of the order of a few days, which is comparable to the temperature changes seen during quiescent accretion flares in e.g., \xte\ \citep[][]{fridriksson2011}, and \maxisource\ \citep[][]{homan2014}. The similar, short decay times in these sources could perhaps be caused by draining (part of) the accretion disc on a viscous time-scale, rather than cooling of the neutron star crust \citep[][]{fridriksson2011,armas2013_2,campana2014,homan2014}. However, \citet{medin2015} suggested that rapid cooling within $\simeq$1 week after the end of an accretion outburst can be explained as cooling of the outermost (fluid) layers of the neutron star called the `ocean'. These authors show that this model can explain the rapid temperature evolution seen in \rx\ \citep[][]{degenaar2013_xtej1709}. However, the decay in the very-faint LMXB \igrmontse\ was strikingly similar to that of \rx, despite a factor of $\simeq$100 difference in outburst accretion rate. A more prominent difference in temperature evolution might therefore be expected, perhaps arguing against the cooling envelope interpretation \citep[][]{armas2013_2}. The observed cooling time-scale for \source, on the other hand, is much longer than the rapid temperature decays seen in these sources, and is fully consistent with crust cooling.

\subsection{Shallow crustal heating}\label{subsec:extraheat}
The occurrence of shallow heating in the crusts of neutron stars, not accounted for by standard nuclear reactions, remains a puzzle. An additional heat source of $\simeq$1--2~MeV is required to properly model the crust-cooling curves of \mxb, \igr, and \exo\ \citep[][]{brown08,degenaar2011_terzan5_3,page2013,degenaar2014_exo3}, whereas a lower value of $\simeq$0.1--0.5~MeV is sufficient for \ks\ and \xte\ (\cite{brown08}; \cite{page2013}, but see \cite{turlione2013}).\footnote{We note that the crust-cooling curve of \maxisource\ has not been modelled yet, but the very high temperatures observed for this source \citep[][]{homan2014} will likely also require the addition of substantial shallow heating.} Furthermore, a number of neutron stars that display superbursts and frequency drifts in mHz quasi-periodic oscillations also appear to require additional shallow heating \citep[e.g.,][]{cumming06,keek2008_1608,keek2009,altamirano2008,altamirano2012,linares2012_ter5_2}.

It is currently unknown whether shallow heating occurs for all neutron stars and, if so, whether the magnitude of the extra energy release should always be the same. The small number of sources and limited constraints on shallow heating in the current sample of crust-cooling curves do not allow us to determine whether the lack/requirement of extra heat is related to source-specific properties such as e.g., mass, spin, magnetic field strength or the type of companion star. One mechanism that could perhaps account for additional heat flux in shallow crustal layers is compositionally-driven convection \citep[][]{medin2011,medin2015,degenaar2013_xtej1709}. This process strongly depends on the composition of the liquid ocean (the remains of nuclear burning on the surface of the neutron star). Therefore, it may give rise to differences in shallow heating for different sources, and a single source may also experience different levels of heating from one outburst to another \citep[][]{medin2015}. This could potentially be tested with crust-cooling observations, e.g., by investigating if the temperature curves of different sources can all be explained with the same amount of shallow heating, or by comparing the crust-cooling curves of a single source after different outbursts.

The crust-cooling curve of \source\ can be adequately modelled using standard physics input, without the need to invoke an additional source of shallow heat. However, a heat source up to $\simeq$1.4~MeV~nucleon$^{-1}$ is still be compatible with the present data (at 2$\sigma$ confidence), which is not unlike that found for \mxb, \igr, and \exo. 

An important caveat is that our conclusions on shallow heating in \source\ are based on the assumption that the neutron star has (nearly) settled at a base temperature similar to that observed prior to the 2012 outburst (i.e., implying that the crust has fully cooled). However, the temperature of the neutron star does not necessarily have to level off at exactly the same value as observed before the outburst. The amount of unburned H/He on the surface of a neutron star determines the heat flux that flows from the crust. Theoretically, for the same core temperature, the observed surface temperature could therefore differ by a factor of a few after different outbursts \citep[][]{brown2002,medin2015}, although there is no strong evidence that this effect is indeed observable. Empirical (power law) decay fits allow for the possibility of continued crust cooling in \source, which would change the derived constraints on the shallow heating. Further observations of Terzan 5 would allow us to confirm/reject that the crust continues to cool, and hence to solidify the constraints on shallow heating in this neutron star. 

\section*{Acknowledgements}
ND acknowledges support via an EU Marie Curie Intra-European fellowship under contract no. FP-PEOPLE-2013-IEF-627148. AC is supported by an NSERC Discovery Grant and is an Associate Member of the CIFAR Cosmology and Gravity programme. COH and GRS are supported by NSERC Discovery Grants, and COH also by an Ingenuity New Faculty Award and Alexander von Humboldt Fellowship. DA acknowledges support from the Royal Society. EFB is supported by NSF grant AST 11-09176. JH acknowledges support provided by the National Aeronautics and Space Administration through Chandra Award Number GO3-14031X  issued by the Chandra X-ray Observatory Center, which is operated by the Smithsonian Astrophysical Observatory for and on behalf of the National Aeronautics Space Administration under contract NAS8-03060. The authors are grateful for the hospitality of the International Space Science Institute (ISSI) in Bern, Switzerland, where part of this work was carried out. 

\footnotesize{
\bibliographystyle{mn2e}

\begin{thebibliography}{89}
\providecommand{\natexlab}[1]{#1}

\bibitem[{{Altamirano} et~al.(2008){Altamirano}, {van der Klis}, {Wijnands} \&
  {Cumming}}]{altamirano2008}
{Altamirano} D., {van der Klis} M., {Wijnands} R., {Cumming} A., 2008, \apjl,
  673, L35

\bibitem[{{Altamirano} et~al.(2015){Altamirano}, {Krimm}, {Patruno},
  {Bahramian}, {Heinke}, {Wijnands} \& {Degenaar}}]{altamirano2015_ter5}
{Altamirano} D., {Krimm} H.~A., {Patruno} A., {Bahramian} A., {Heinke} C.~O.,
  {Wijnands} R., {Degenaar} N., 2015, \atel, 7240

\bibitem[{{Altamirano} et~al.(2012)}]{altamirano2012}
{Altamirano} D. et~al., 2012, \mnras, 426, 927

\bibitem[{{Armas Padilla} et~al.(2013){Armas Padilla}, {Wijnands} \&
  {Degenaar}}]{armas2013_2}
{Armas Padilla} M., {Wijnands} R., {Degenaar} N., 2013, \mnras, 436, L89

\bibitem[{{Arnaud}(1996)}]{xspec}
{Arnaud} K., 1996, in G.~{Jacoby}, J.~{Barnes}, eds, ASP Conf. Ser. Vol. 101, Astronomical Data Analysis
  Software and Systems V. Astron. Soc. Pac., San Francisco, p. 17

\bibitem[{{Bahramian} et~al.(2014)}]{bahramian2014}
{Bahramian} A. et~al., 2014, \apj, 780, 127

\bibitem[{{Barret}(2012)}]{barret2012}
{Barret} D., 2012, \apj, 753, 84

\bibitem[{{Bernardini} et~al.(2013)}]{bernardini2013}
{Bernardini} F., {Cackett} E., {Brown} E., {D'Angelo} C., {Degenaar} N.,
  {Miller} J., {Reynolds} M., {Wijnands} R., 2013, \mnras, 436, 2465

\bibitem[{{Brown} \& {Cumming}(2009)}]{brown08}
{Brown} E., {Cumming} A., 2009, \apj, 698, 1020

\bibitem[{{Brown} et~al.(1998){Brown}, {Bildsten} \& {Rutledge}}]{brown1998}
{Brown} E., {Bildsten} L., {Rutledge} R., 1998, \apjl, 504, L95

\bibitem[{{Brown} et~al.(2002){Brown}, {Bildsten} \& {Chang}}]{brown2002}
{Brown} E., {Bildsten} L., {Chang} P., 2002, \apj, 574, 920

\bibitem[{{Cackett} et~al.(2006){Cackett}, {Wijnands}, {Linares}, {Miller},
  {Homan} \& {Lewin}}]{cackett2006}
{Cackett} E., {Wijnands} R., {Linares} M., {Miller} J., {Homan} J., {Lewin} W.,
  2006, \mnras, 372, 479

\bibitem[{{Cackett} et~al.(2008){Cackett}, {Wijnands}, {Miller}, {Brown} \&
  {Degenaar}}]{cackett2008}
{Cackett} E., {Wijnands} R., {Miller} J., {Brown} E., {Degenaar} N., 2008,
  \apjl, 687, L87

\bibitem[{{Cackett} et~al.(2010{\natexlab{a}}){Cackett}, {Brown}, {Miller} \&
  {Wijnands}}]{cackett2010_cenx4}
{Cackett} E., {Brown} E., {Miller} J., {Wijnands} R., 2010{\natexlab{a}}, \apj,
  720, 1325

\bibitem[{{Cackett} et~al.(2010{\natexlab{b}}){Cackett}, {Brown}, {Cumming},
  {Degenaar}, {Miller} \& {Wijnands}}]{cackett2010}
{Cackett} E., {Brown} E., {Cumming} A., {Degenaar} N., {Miller} J., {Wijnands}
  R., 2010{\natexlab{b}}, \apjl, 722, L137

\bibitem[{{Cackett} et~al.(2011){Cackett}, {Fridriksson}, {Homan}, {Miller} \&
  {Wijnands}}]{cackett2011_aqlx1}
{Cackett} E., {Fridriksson} J., {Homan} J., {Miller} J., {Wijnands} R., 2011,
  \mnras, 414, 3006

\bibitem[{{Cackett} et~al.(2013)}]{cackett2013_1659}
{Cackett} E., {Brown} E., {Cumming} A., {Degenaar} N., {Fridriksson} J.,
  {Homan} J., {Miller} J., {Wijnands} R., 2013, \apj, 774, 131

\bibitem[{{Campana} et~al.(1998){Campana}, {Colpi}, {Mereghetti}, {Stella} \&
  {Tavani}}]{campana1998}
{Campana} S., {Colpi} M., {Mereghetti} S., {Stella} L., {Tavani} M., 1998,
  A\&AR, 8, 279

\bibitem[{{Campana} et~al.(2013){Campana}, {Coti Zelati} \&
  {D'Avanzo}}]{campana2013}
{Campana} S., {Coti Zelati} F., {D'Avanzo} P., 2013, \mnras, 432, 1695

\bibitem[{{Campana} et~al.(2014)}]{campana2014}
{Campana} S., {Brivio} F., {Degenaar} N., {Mereghetti} S., {Wijnands} R.,
  {D'Avanzo} P., {Israel} G.~L., {Stella} L., 2014, \mnras, 441, 1984

\bibitem[{{Chakrabarty} et~al.(2014)}]{chakrabarty2014_cenx4}
{Chakrabarty} D. et~al., 2014, \apj, 797, 92

\bibitem[{{Cohn} et~al.(2002){Cohn}, {Lugger}, {Grindlay} \&
  {Edmonds}}]{cohn2002}
{Cohn} H., {Lugger} P., {Grindlay} J., {Edmonds} P., 2002, \apj, 571, 818

\bibitem[{{Cominsky} \& {Wood}(1984)}]{cominsky1984}
{Cominsky} L.~R., {Wood} K.~S., 1984, \apj, 283, 765

\bibitem[{{Cornelisse} et~al.(2012)}]{cornelisse2012}
{Cornelisse} R. et~al., 2012, \mnras, 420, 3538

\bibitem[{{Cumming} et~al.(2006){Cumming}, {Macbeth}, {in 't Zand} \&
  {Page}}]{cumming06}
{Cumming} A., {Macbeth} J., {in 't Zand} J., {Page} D., 2006, \apj, 646, 429

\bibitem[{{Degenaar} \& {Wijnands}(2011{\natexlab{a}})}]{degenaar2011_terzan5}
{Degenaar} N., {Wijnands} R., 2011{\natexlab{a}}, \mnras, 412, L68

\bibitem[{{Degenaar} \&
  {Wijnands}(2011{\natexlab{b}})}]{degenaar2011_terzan5_2}
{Degenaar} N., {Wijnands} R., 2011{\natexlab{b}}, \mnras, 414, L50

\bibitem[{{Degenaar} \& {Wijnands}(2012)}]{degenaar2012_1745}
{Degenaar} N., {Wijnands} R., 2012, \mnras, 422, 581

\bibitem[{{Degenaar} et~al.(2009)}]{degenaar09_exo1}
{Degenaar} N. et~al., 2009, \mnras, 396, L26

\bibitem[{{Degenaar} et~al.(2011{\natexlab{a}})}]{degenaar2010_exo2}
{Degenaar} N. et~al., 2011{\natexlab{a}}, \mnras, 412, 1409

\bibitem[{{Degenaar} et~al.(2011{\natexlab{b}}){Degenaar}, {Brown} \&
  {Wijnands}}]{degenaar2011_terzan5_3}
{Degenaar} N., {Brown} E., {Wijnands} R., 2011{\natexlab{b}}, \mnras, 418, L152

\bibitem[{{Degenaar} et~al.(2012){Degenaar}, {Patruno} \&
  {Wijnands}}]{degenaar2012_amxp}
{Degenaar} N., {Patruno} A., {Wijnands} R., 2012, \apj, 756, 148

\bibitem[{{Degenaar} et~al.(2013{\natexlab{a}}){Degenaar}, {Wijnands} \&
  {Miller}}]{degenaar2013_xtej1709}
{Degenaar} N., {Wijnands} R., {Miller} J., 2013{\natexlab{a}}, \apjl, 767, L31

\bibitem[{{Degenaar} et~al.(2013{\natexlab{b}})}]{degenaar2013_terzan5}
{Degenaar} N. et~al., 2013{\natexlab{b}}, \apj, 775, 48

\bibitem[{{Degenaar} et~al.(2014)}]{degenaar2014_exo3}
{Degenaar} N. et~al., 2014, \apj, 791, 47

\bibitem[{{D{\'{\i}}az Trigo} et~al.(2011){D{\'{\i}}az Trigo}, {Boirin},
  {Costantini}, {M{\'e}ndez} \& {Parmar}}]{diaztrigo2011}
{D{\'{\i}}az Trigo} M., {Boirin} L., {Costantini} E., {M{\'e}ndez} M., {Parmar}
  A., 2011, \aap, 528, 150

\bibitem[{{Estrad{\'e}} et~al.(2011)}]{estrade2011}
{Estrad{\'e}} A. et~al., 2011, Physical Review Letters, 107, 172503

\bibitem[{{Fridriksson} et~al.(2010)}]{fridriksson2010}
{Fridriksson} J. et~al., 2010, \apj, 714, 270

\bibitem[{{Fridriksson} et~al.(2011)}]{fridriksson2011}
{Fridriksson} J. et~al., 2011, \apj, 736, 162

\bibitem[{{Galloway} et~al.(2008){Galloway}, {Muno}, {Hartman}, {Psaltis} \&
  {Chakrabarty}}]{galloway06}
{Galloway} D., {Muno} M., {Hartman} J., {Psaltis} D., {Chakrabarty} D., 2008,
  \apjs, 179, 360

\bibitem[{{Galloway} et~al.(2010){Galloway}, {Lin}, {Chakrabarty} \&
  {Hartman}}]{galloway2010}
{Galloway} D., {Lin} J., {Chakrabarty} D., {Hartman} J., 2010, \apjl, 711, L148

\bibitem[{{Haensel} \& {Zdunik}(1990)}]{haensel1990a}
{Haensel} P., {Zdunik} J., 1990, \aap, 227, 431

\bibitem[{{Haensel} \& {Zdunik}(2008)}]{haensel2008}
{Haensel} P., {Zdunik} J., 2008, \aap, 480, 459

\bibitem[{{Heinke} et~al.(2003){Heinke}, {Edmonds}, {Grindlay}, {Lloyd}, {Cohn}
  \& {Lugger}}]{heinke2003}
{Heinke} C., {Edmonds} P., {Grindlay} J., {Lloyd} D., {Cohn} H., {Lugger} P.,
  2003, \apj, 590, 809

\bibitem[{{Heinke} et~al.(2006{\natexlab{a}}){Heinke}, {Rybicki}, {Narayan} \&
  {Grindlay}}]{heinke2006}
{Heinke} C., {Rybicki} G., {Narayan} R., {Grindlay} J., 2006{\natexlab{a}},
  \apj, 644, 1090

\bibitem[{{Heinke} et~al.(2006{\natexlab{b}}){Heinke}, {Wijnands}, {Cohn},
  {Lugger}, {Grindlay}, {Pooley} \& {Lewin}}]{heinke2006_terzan5}
{Heinke} C., {Wijnands} R., {Cohn} H., {Lugger} P., {Grindlay} J., {Pooley} D.,
  {Lewin} W., 2006{\natexlab{b}}, \apj, 651, 1098

\bibitem[{{Homan} et~al.(2014)}]{homan2014}
{Homan} J., {Fridriksson} J.~K., {Wijnands} R., {Cackett} E.~M., {Degenaar} N.,
  {Linares} M., {Lin} D., {Remillard} R.~A., 2014, \apj, 795, 131

\bibitem[{{Horowitz} et~al.(2008){Horowitz}, {Dussan} \&
  {Berry}}]{horowitz2008}
{Horowitz} C., {Dussan} H., {Berry} D., 2008, \prc, 77, 045807

\bibitem[{{Horowitz} et~al.(2015){Horowitz}, {Berry}, {Briggs}, {Caplan},
  {Cumming} \& {Schneider}}]{horowitz2015}
{Horowitz} C.~J., {Berry} D.~K., {Briggs} C.~M., {Caplan} M.~E., {Cumming} A.,
  {Schneider} A.~S., 2015, Physical Review Letters, 114, 031102

\bibitem[{{in 't Zand} et~al.(2007){in 't Zand}, {Jonker} \&
  {Markwardt}}]{zand07}
{in 't Zand} J., {Jonker} P., {Markwardt} C., 2007, \aap, 465, 953

\bibitem[{{Jonker} et~al.(2004){Jonker}, {Galloway}, {McClintock}, {Buxton},
  {Garcia} \& {Murray}}]{jonker2004}
{Jonker} P., {Galloway} D., {McClintock} J., {Buxton} M., {Garcia} M., {Murray}
  S., 2004, \mnras, 354, 666

\bibitem[{{Keek} et~al.(2008){Keek}, {in 't Zand}, {Kuulkers}, {Cumming},
  {Brown} \& {Suzuki}}]{keek2008_1608}
{Keek} L., {in 't Zand} J., {Kuulkers} E., {Cumming} A., {Brown} E., {Suzuki}
  M., 2008, \aap, 479, 177

\bibitem[{{Keek} et~al.(2009){Keek}, {Langer} \& {in 't Zand}}]{keek2009}
{Keek} L., {Langer} N., {in 't Zand} J., 2009, \aap, 502, 871

\bibitem[{{Lattimer} et~al.(1994){Lattimer}, {van Riper}, {Prakash} \&
  {Prakash}}]{lattimer1994}
{Lattimer} J., {van Riper} K., {Prakash} M., {Prakash} M., 1994, \apj, 425, 802

\bibitem[{{Lin} et~al.(2009){Lin}, {Remillard} \& {Homan}}]{lin2009}
{Lin} D., {Remillard} R., {Homan} J., 2009, \apj, 696, 1257

\bibitem[{{Linares} et~al.(2012){Linares}, {Altamirano}, {Chakrabarty},
  {Cumming} \& {Keek}}]{linares2012_ter5_2}
{Linares} M., {Altamirano} D., {Chakrabarty} D., {Cumming} A., {Keek} L., 2012,
  \apj, 748, 82

\bibitem[{{Medin} \& {Cumming}(2011)}]{medin2011}
{Medin} Z., {Cumming} A., 2011, \apj, 730, 97

\bibitem[{{Medin} \& {Cumming}(2015)}]{medin2015}
{Medin} Z., {Cumming} A., 2015, \apj, 802, 29

\bibitem[{{Miller} et~al.(2009){Miller}, {Cackett} \& {Reis}}]{miller2009}
{Miller} J., {Cackett} E., {Reis} R., 2009, \apjl, 707, L77

\bibitem[{{Miller} et~al.(2011){Miller}, {Maitra}, {Cackett}, {Bhattacharyya}
  \& {Strohmayer}}]{miller2011}
{Miller} J., {Maitra} D., {Cackett} E., {Bhattacharyya} S., {Strohmayer} T.,
  2011, \apjl, 731, L7

\bibitem[{{Miller}(2013)}]{miller2013_NSreview}
{Miller} M.~C., 2013, preprint (arXiv:1312.0029)

\bibitem[{{Motta} et~al.(2011)}]{motta2011}
{Motta} S. et~al., 2011, \mnras, 414, 1508

\bibitem[{{Muno} et~al.(2000){Muno}, {Fox}, {Morgan} \& {Bildsten}}]{muno2000}
{Muno} M.~P., {Fox} D.~W., {Morgan} E.~H., {Bildsten} L., 2000, \apj, 542, 1016

\bibitem[{{Muno} et~al.(2001){Muno}, {Chakrabarty}, {Galloway} \&
  {Savov}}]{muno2001}
{Muno} M.~P., {Chakrabarty} D., {Galloway} D.~K., {Savov} P., 2001, \apjl, 553,
  L157

\bibitem[{{Ortolani} et~al.(2007){Ortolani}, {Barbuy}, {Bica}, {Zoccali} \&
  {Renzini}}]{ortolani2007}
{Ortolani} S., {Barbuy} B., {Bica} E., {Zoccali} M., {Renzini} A., 2007, \aap,
  470, 1043

\bibitem[{{{\"O}zel}(2013)}]{ozel2013_NSreview}
{{\"O}zel} F., 2013, Rep. Prog. Phys., 76, 016901

\bibitem[{{Page} \& {Reddy}(2012)}]{page2012}
{Page} D., {Reddy} S., 2012, ``Neutron Star Crust", ed. Bertulani,
  C. A. \& Piekarewicz, J., Nova Science Publishers, Hauppauge, preprint (arXiv:1201.5602)

\bibitem[{{Page} \& {Reddy}(2013)}]{page2013}
{Page} D., {Reddy} S., 2013, Physical Review Letters, 111, 241102

\bibitem[{{Papitto} et~al.(2011)}]{papitto2010}
{Papitto} A., {D'A{\`i}} A., {Motta} S., {Riggio} A., {Burderi} L., {di Salvo}
  T., {Belloni} T., {Iaria} R., 2011, \aap, 526, L3

\bibitem[{{Parmar} et~al.(1986){Parmar}, {White}, {Giommi} \&
  {Gottwald}}]{parmar1986}
{Parmar} A., {White} N., {Giommi} P., {Gottwald} M., 1986, \apj, 308, 199

\bibitem[{{Revnivtsev} et~al.(2001){Revnivtsev}, {Churazov}, {Gilfanov} \&
  {Sunyaev}}]{revnivtsev2001}
{Revnivtsev} M., {Churazov} E., {Gilfanov} M., {Sunyaev} R., 2001, \aap, 372,
  138

\bibitem[{{Rutledge} et~al.(2002{\natexlab{a}}){Rutledge}, {Bildsten}, {Brown},
  {Pavlov} \& {Zavlin}}]{rutledge2002_aqlX1}
{Rutledge} R., {Bildsten} L., {Brown} E., {Pavlov} G., {Zavlin} V.,
  2002{\natexlab{a}}, \apj, 577, 346

\bibitem[{{Rutledge} et~al.(2002{\natexlab{b}}){Rutledge}, {Bildsten}, {Brown},
  {Pavlov}, {Zavlin} \& {Ushomirsky}}]{rutledge2002}
{Rutledge} R., {Bildsten} L., {Brown} E., {Pavlov} G., {Zavlin} V.,
  {Ushomirsky} G., 2002{\natexlab{b}}, \apj, 580, 413

\bibitem[{{Serino} et~al.(2012)}]{serino2012}
{Serino} M., {Mihara} T., {Matsuoka} M., {Nakahira} S., {Sugizaki} M., {Ueda}
  Y., {Kawai} N., {Ueno} S., 2012, \pasj, 64, 91

\bibitem[{{Shternin} et~al.(2007){Shternin}, {Yakovlev}, {Haensel} \&
  {Potekhin}}]{shternin07}
{Shternin} P., {Yakovlev} D., {Haensel} P., {Potekhin} A., 2007, \mnras, 382,
  L43

\bibitem[{{Smith} et~al.(1997){Smith}, {Morgan} \& {Bradt}}]{smith1997}
{Smith} D., {Morgan} E., {Bradt} H., 1997, \apjl, 479, L137

\bibitem[{{Steiner}(2012)}]{steiner2012}
{Steiner} A., 2012, \prc, 85, 055804

\bibitem[{{Tremou} et~al.(2015)}]{tremou2015}
{Tremou} E. et~al., 2015, \atel, 7262

\bibitem[{{Turlione} et~al.(2015){Turlione}, {Aguilera} \&
  {Pons}}]{turlione2013}
{Turlione} A., {Aguilera} D., {Pons} J., 2015, \aap, 577, 5

\bibitem[{{Verner} et~al.(1996){Verner}, {Ferland}, {Korista} \&
  {Yakovlev}}]{verner1996}
{Verner} D., {Ferland} G., {Korista} K., {Yakovlev} D., 1996, \apj, 465, 487

\bibitem[{{Wijnands} et~al.(2001{\natexlab{a}}){Wijnands}, {Strohmayer} \&
  {Franco}}]{wijnands01}
{Wijnands} R., {Strohmayer} T., {Franco} L., 2001{\natexlab{a}}, \apjl, 549,
  L71

\bibitem[{{Wijnands} et~al.(2001{\natexlab{b}}){Wijnands}, {Miller},
  {Markwardt}, {Lewin} \& {van der Klis}}]{wijnands2001}
{Wijnands} R., {Miller} J., {Markwardt} C., {Lewin} W., {van der Klis} M.,
  2001{\natexlab{b}}, \apjl, 560, L159

\bibitem[{{Wijnands} et~al.(2002){Wijnands}, {Guainazzi}, {van der Klis} \&
  {M{\'e}ndez}}]{wijnands2002}
{Wijnands} R., {Guainazzi} M., {van der Klis} M., {M{\'e}ndez} M., 2002, \apjl,
  573, L45

\bibitem[{{Wijnands} et~al.(2003){Wijnands}, {Nowak}, {Miller}, {Homan},
  {Wachter} \& {Lewin}}]{wijnands2003}
{Wijnands} R., {Nowak} M., {Miller} J., {Homan} J., {Wachter} S., {Lewin} W.,
  2003, \apj, 594, 952

\bibitem[{{Wijnands} et~al.(2004){Wijnands}, {Homan}, {Miller} \&
  {Lewin}}]{wijnands2004}
{Wijnands} R., {Homan} J., {Miller} J., {Lewin} W., 2004, \apjl, 606, L61

\bibitem[{{Wijnands} et~al.(2005){Wijnands}, {Homan}, {Heinke}, {Miller} \&
  {Lewin}}]{wijnands05_amxps}
{Wijnands} R., {Homan} J., {Heinke} C., {Miller} J., {Lewin} W., 2005, \apj,
  619, 492

\bibitem[{{Wijnands} et~al.(2013){Wijnands}, {Degenaar} \&
  {Page}}]{wijnands2012}
{Wijnands} R., {Degenaar} N., {Page} D., 2013, \mnras, 432, 2366

\bibitem[{{Wilms} et~al.(2000){Wilms}, {Allen} \& {McCray}}]{wilms2000}
{Wilms} J., {Allen} A., {McCray} R., 2000, \apj, 542, 914

\end{thebibliography}

}

\end{document}